\documentclass[preprintnumbers,amsmath,amssymb,
12pt,floatfix,superscriptaddress,nofootinbib]{revtex4}
\usepackage{graphicx}
\usepackage{epsfig}
\usepackage{bm}
\usepackage{amsfonts}
\def\be {\begin{equation}}
\def\ee {\end{equation}}
\begin{document}

\title{Can a wormhole generate electromagnetic field?}

\author{Mubasher Jamil}
\affiliation{Center for Advanced Mathematics and Physics, National
University of Sciences and Technology,\\ Rawalpindi, 46000,
Pakistan}

\begin{abstract}
We have considered the possibility of a rotating wormhole surrounded
by a cloud of charged particles. Due to slow rotation of the
wormhole, the charged particles are dragged, thereby producing an
electromagnetic field. We have determined the strength of this
electromagnetic field and the corresponding flux of radiation.
\end{abstract}
\maketitle
\newpage
\section{Introduction}
Traversable wormhole arises as a solution to the Einstein field
equations and was first proposed by Morris and Thorne
\cite{morris,morris1} as time travel machines. The idea of wormhole
spacetime was given by J.A. Wheeler in his attempt to apply quantum
mechanics at the Planck scale. The resulting spacetime turns out to
be fluctuating giving rise a number of topologies including the
wormhole \cite{wheeler}. A static and spherically symmetric wormhole
possesses interesting geometry having a throat that flares out in
two opposite directions. The throat connects either two different
asymptotically flat regions in the same spacetime or entirely two
distinct spacetimes. Later on, Ellis \cite{ellis} termed this
geometry as a `drainhole' that could render particle motion from
either mouths.  The throat has the tendency to get closed in a very
short time (of the order of Planck time) thereby limiting the time
travel possibility. In order to create a stable wormhole, a negative
energy (or the exotic matter) is required to keep the wormhole's
throat open. Such a negative energy thus violates the \textit{null
energy condition} (NEC) i.e. $T_{\mu\nu}u^{\mu}u^{\nu}\geq0$, where
$T_{\mu\nu}$ is the stress energy tensor and $u^{\mu}$ is the future
directed null vector. Since NEC is the weakest energy condition, it
implies that all the energy conditions (weak, strong and dominant)
will be violated automatically. These energy conditions are
generally obeyed by the classical matter but are violated by certain
quantum fields which exhibit the Casimir effect and the Hawking
evaporation process \cite{lobo3}.

In the context of cosmology, the phantom like dark energy with the
equation of state (EoS) $\omega<-1$ violates the NEC and has been
analyzed as a source to sustain traversable wormholes
\cite{lobo1,lobo2}. It has been shown that arbitrary small amount of
phantom energy can support the existence of wormhole
\cite{rahman,rahaman1,rahaman2,peter}. It is argued in
\cite{hayward} that traversable wormhole can be constructed from the
black hole by absorbing exotic energy and conversely the wormhole
can collapse to a black hole by releasing exotic energy.
Surprisingly, the wormhole can lead to inflationary universe by
absorbing arbitrarily large amount of exotic matter. Moreover, the
size of the wormhole can be increased or decreased by a
corresponding increase or decrease of the absorption of exotic
matter \cite{jamil}.

Earlier, it has been shown that neutral test particles propagating
towards the rotating wormhole radially, start moving about the
wormhole in the spiral path. After entering the wormhole's throat,
the particles pass through the throat and move away from the throat
following the spiral trajectories \cite{sushkov}. We are here
interested in a stationary and axially symmetric wormhole having
non-zero angular velocity surrounded by the continuum of charged
particles that are dragged by the wormhole in the angular direction.
The rotation is assumed to be negligible so that quadratic terms in
the angular velocity are ignored. The frame dragging effects on the
charged particles produces a poloidal electromagnetic field. The
resulting field around the wormhole is determined under the slow
rotation approximation. This feature of the wormhole physics can be
of interest from astronomical point of view. It has been suggested
that wormholes can cause gravitational lensing events \cite{dey}.
Also the accretion of matter on a wormhole could also create a black
hole \cite{kardashev}. In this connection, the electromagnetic
structure of rotating wormholes is investigated here.

The outline of this paper is as follows: In the next section, we
formulate the governing equations of our gravitational system and
then solve them in the third section. Finally, we conclude in the
fourth section.

\section{Formulation of the dynamical system}

We consider a stationary and axially symmetric wormhole, surrounded
by the charged particles, is given by (in geometrical units $G=1=c$)
\cite{lobo3}:
\begin{equation}
ds^2=-N^2dt^2+e^\mu dr^2+r^2K^2[d\theta^2+\sin^2\theta(d\phi-\omega
dt)^2],
\end{equation}
where
\begin{equation}
e^{-\mu(r,\theta)}=1-\frac{b(r,\theta)}{r},
\end{equation}
with $N$, $K$, $\mu$ and $\omega$ are functions of $r$ and $\theta$
only. The function $b(r,\theta)$ determines the spatial shape of the
wormhole and is thus called the \textit{shape function}. Also
$\omega(r,\theta)$ is the orbital angular velocity of the wormhole.
More precisely
\begin{equation}
\omega\equiv\frac{d\phi}{dt}=\frac{d\phi/d\tau}{dt/d\tau}=\frac{u^\phi}{u^t},
\end{equation}
where $u^\phi$ and $u^t$ are azimuthal and transverse components of
the four velocity respectively. Notice that static and spherically
symmetric Morris-Thorne wormhole \cite{morris} is obtained from Eq.
(1) by doing the following substitution
\begin{equation}
N(r,\theta)=e^{2\Phi(r)}, \ \ b(r,\theta)=b(r),\ \ K(r,\theta)=1,\ \
\omega(r,\theta)=0.
\end{equation}
Above the \textit{redshift function} $\Phi(r)$ determines the
gravitational redshift. In the slow rotation approximation
$O(\omega^2)$, Eq. (1) gives
\begin{equation}
{ds}^2\approx-N^2dt^2+e^\mu
dr^2+r^2K^2[d\theta^2+\sin^2\theta(d\phi^2-2\omega dtd\phi)].
\end{equation}
The determinant of the metric Eq. (5) is
\begin{equation}
|g_{\alpha\beta}|=g=-e^\mu r^4K^4N^2\sin^2\theta,
\end{equation}
where we have neglected terms in $O(\omega^2)$. The relativistic
Maxwell equations are
\begin{equation}
F_{[\alpha\beta,\gamma]}=0,
\end{equation}
\begin{equation}
(\sqrt{-g}F^{\alpha\beta})_{,\beta}=4\pi J^\alpha.
\end{equation}
Above $F_{\alpha\beta}$ is the electromagnetic field tensor and
$J^\alpha$ is the four vector current density of the charged
particles. The former is given by \cite{andre}
\begin{equation}
F_{\alpha\beta}=u_\alpha E_\beta-u_\beta
E_\alpha+\eta_{\alpha\beta\gamma\delta}u^\gamma B^\delta.
\end{equation}
Here $u^\alpha$ is the four velocity of the charged particles
producing an electromagnetic field having $E^\alpha$ and $B^\alpha$
as four vector electric and magnetic field components. The current
density containing the convection and the conduction currents is
given by \cite{andre}
\begin{equation}
J^\alpha=\epsilon u^\alpha+\sigma u_\beta F^{\beta\alpha}.
\end{equation}
Here $\epsilon$ and $\sigma$ are the charge density and the fluid
conductivity respectively. Moreover, the volume four element is
given by \cite{landau}
\begin{equation}
\eta_{\alpha\beta\gamma\delta}=\sqrt{-g}\epsilon_{\alpha\beta\gamma\delta},\
\
\eta^{\alpha\beta\gamma\delta}=-\frac{1}{\sqrt{-g}}\epsilon^{\alpha\beta\gamma\delta}.
\end{equation}
Above $\epsilon_{\alpha\beta\gamma\delta}$ is the Levi-Civita
symbol. We now consider a zero angular momentum observer (ZAMO)
moving about the slowly rotating wormhole in the equatorial plane
($\theta=\pi/2$) at a fixed distance from the wormhole's axis of
rotation ($r=R$) which yield $u^r=0=u^\theta$. The four velocity of
ZAMO is given by
\begin{equation}
u^{\alpha}=N^{-1}(r,\theta)(1,0,0,\omega(r,\theta)),\
u_{\alpha}=N(r,\theta)(-1,0,0,0),
\end{equation}
which satisfy the normalization condition $u^\alpha u_\alpha=-1$.
Thus the electromagnetic field around the wormhole is determined
from Eqs. (7) and (8).

Assuming the neutrality of plasma ($J^{\alpha}=0$) and the
electromagnetic field to be function of $r$ and $\theta$ only,
equations (7) for the metric in Eq. (5) become
\begin{widetext}
\begin{eqnarray}
(r^2e^{\mu/2}K^2\sin\theta B^r)_{,r}+(r^2e^{\mu/2}K^2\sin\theta
B^\theta)_{,\theta}&=&0,\\
 (NE_\phi)_{,r}&=&0,\\
  (NE_\phi)_{,\theta}&=&0,\\
(NE_\theta+\omega r^2e^{\mu/2}K^2\sin\theta B^r)_{,r}-(NE_r+\omega
r^2e^{\mu/2}K^2\sin\theta B^\theta)_{,\theta}&=&0,
\end{eqnarray}
\end{widetext}
while Eq. (8) yields
\begin{widetext}
\begin{eqnarray}
(r^2e^{\mu/2}K^2\sin\theta E^r)_{,r}+(r^2e^{\mu/2}K^2\sin\theta
E^\theta)_{,\theta}&=&0,\\
(NB_\phi)_{,r}&=&0,\\
(NB_\phi)_{,\theta}&=&0,\\
(NB_\theta+\omega r^2e^{\mu/2}K^2\sin\theta E^r)_{,r}-(NB_r+\omega
r^2e^{\mu/2}K^2\sin\theta E^\theta)_{,\theta}&=&0.
\end{eqnarray}
\end{widetext}
\section{Solution of governing equations}

To obtain some specific solution of Eqs. (7) and (8), let us assume
the following ansatz for the electric and the magnetic fields:
\begin{equation}
E_r(r,\theta)\equiv k_1E_\theta(r,\theta)\equiv
k_2E_\phi(r,\theta)=R_E(r)\Theta_E(\theta),\\
\end{equation}
\begin{equation}
B_r(r,\theta)\equiv k_3B_\theta(r,\theta)\equiv
k_4B_\phi(r,\theta)=R_B(r)\Theta_B(\theta).
\end{equation}
Here $k_i$, $i=1,2,3,4$ are non-zero dimensional constants. We
consider two types of solutions containing either (1) the arbitrary
functions $N$, $K$, $\mu$ and $\omega$ are functions of $r$ only or
(2) the same arbitrary functions are dependent on $\theta$ only.

Case (1). We first assume the arbitrary functions in the Eq. (5) to
be dependent on the radial coordinate $r$ only. In the following,
the $C_j$ where $j=1,..,12$ and $D_l$ where $l=1,..,9$ are constants
parameters.

Using Eq. (22) in (13), we get
\begin{equation}
R_B(r)=\frac{D_1e^{C_1r}}{r^2e^{\mu(r)/2}K^2(r)},\ \
\Theta_B(\theta)=\frac{D_2e^{-C_1k_3\theta}}{\sin\theta}.
\end{equation}
Above $C_1$ has dimensions of $L^{-1}$ while $k_3$ has dimensions of
$L$. Using Eq. (21) in (14) we get
\begin{equation}
R_E(r)=\frac{C_2}{N(r)}.
\end{equation}
Similarly, using Eq. (21) in (15), we have
\begin{equation}
\Theta_E(\theta)=C_3.
\end{equation}
Finally, using Eqs. (23-25) in (16), we get
\begin{equation}
\omega(r)_{,r}+2C_1\omega(r)=0,
\end{equation}
which gives
\begin{equation}
\omega(r)=D_3e^{-2C_1r}.
\end{equation}
Notice that if $D_3$ and $C_1$ are positive constants then the
angular velocity $\omega$ will decrease as $r$ increases. Making use
of Eq. (21) in (17), we get
\begin{equation}
R_E(r)=\frac{D_4e^{C_4r}}{r^2e^{\mu(r)/2}K^2(r)},\ \
\Theta_E(\theta)=\frac{D_5e^{-C_4k_1\theta}}{\sin\theta}.
\end{equation}
Above $C_4$ has dimensions of $L^{-1}$ while $k_1$ has dimensions of
$L$. Using Eq. (22) in (18) gives
\begin{equation}
R_B(r)=\frac{C_5}{N(r)}.
\end{equation}
Also, Eq. (22) in (19) gives
\begin{equation}
\Theta_B(\theta)=C_6.
\end{equation}

The general solution of the Eqs. (13-20) becomes
\begin{equation}
E_r(r,\theta)=R_E(r)\Theta_E(\theta)=\frac{D_4D_5e^{C_4(r-k_1\theta)}}
{r^2e^{\mu(r)/2}K^2(r)\sin\theta},
\end{equation}
\begin{equation}
B_r(r,\theta)=R_B(r)\Theta_B(\theta)=\frac{D_1D_2e^{C_1(r-k_3\theta)}}
{r^2e^{\mu(r)/2}K^2(r)\sin\theta}.
\end{equation}

Case (2). Let us solve Eqs. (13-20) by choosing the arbitrary
functions in Eq. (5) depending on $\theta$ only.

Using Eq. (22) in (13), we get
\begin{equation}
R_B(r)=\frac{D_{5}e^{C_7r}}{r^2},\ \
\Theta_B(\theta)=\frac{D_{6}e^{-k_3C_7\theta}}{e^{\mu(\theta)/2}K^2(\theta)\sin\theta}.
\end{equation}
Above $C_7$ has dimensions of $L^{-1}$. Using Eq. (21) in (14), we
have
\begin{equation}
R_E(r)=C_{8}.
\end{equation}
Also using Eq. (21) in (15), we can write
\begin{equation}
\Theta_E(\theta)=\frac{C_{9}}{N(\theta)}.
\end{equation}
Using Eqs. (33-35) in (16), we get
\begin{equation}
\omega(\theta)_{,\theta}-2C_7k_{3}\omega(\theta)=0,
\end{equation}
which yields
\begin{equation}
\omega(\theta)=D_7e^{2C_{7}k_3\theta}.
\end{equation}
Further using Eq. (21) in (17), we get
\begin{equation}
R_E(r)=\frac{D_8e^{C_{10}r}}{r^2},\ \
\Theta_E(\theta)=\frac{D_9e^{-k_1C_{10}\theta}}{e^{\mu(\theta)/2}K^2(\theta)\sin\theta}.
\end{equation}
Similarly, $C_{10}$ has dimensions of $L^{-1}$. Making use of Eq.
(22) in (18), we get
\begin{equation}
R_B(r)=C_{11}.
\end{equation}
Using Eq. (22) in (19) yields
\begin{equation}
\Theta_B(\theta)=\frac{C_{12}}{N(\theta)}.
\end{equation}
From Eqs. (32-38), the components of electric and magnetic field
become
\begin{equation}
E_r(r,\theta)=R_E(r)\Theta_E(\theta)=\frac{D_8D_9e^{C_{10}
(r-k_1\theta)}}{r^2e^{\mu(\theta)/2}K^2(\theta)\sin\theta},
\end{equation}
\begin{equation}
B_r(r,\theta)=R_B(r)\Theta_B(\theta)=\frac{D_5D_6e^{C_{7}
(r-k_3\theta)}}{r^2e^{\mu(\theta)/2}K^2(\theta)\sin\theta}.
\end{equation}

The Poynting vector $\mathbf{S}=\mathbf{E\times B}$ determining the
flux of radiation from the slowly rotating wormhole is determined to
be
\begin{equation}
\mathbf{S}=E_{\phi}B_{\phi}\Delta\equiv\frac{1}{k_2k_4}E_rB_r\Delta
\equiv\frac{k_1k_3}{k_2k_4}E_\theta
B_\theta\Delta,
\end{equation}
where
\begin{equation}
\Delta\equiv\hat{i}\left(\frac{k_2}{k_1}-\frac{k_4}{k_3}\right)-\hat{j}
(k_2-k_4)+\hat{k}\left(\frac{1}{k_3}-\frac{1}{k_1}\right)k_2k_4.
\end{equation}

\section{Conclusion}

In this paper we have studied the electromagnetic field and the
corresponding generation of electromagnetic flux due to a rotating
wormhole. The source of this field is the charge density which
surrounds the wormhole. The distribution of charges is assumed to be
spherically symmetrical. In general, a wormhole spacetime is
inhomogeneous and therefore requires inhomogeneous distribution of
matter (or any other source) in its vicinity \cite{jamil1}. But this
does not apply here since exterior spacetime of the wormhole
contains only electric charges which do not induce any pressure on
the wormhole. Also note that metric (1) contains three arbitrary
functions, which need to be chosen for any physically motivated
model. Our model predicts the production of the electromagnetic
field with a certain radiation flux due to wormhole rotation. It is
anticipated that this model would explain some bizarre phenomenon
processes in the universe like the gamma ray bursts. Finally, the
analysis performed in this paper can be extended to charged coupled
with rotating wormholes as well \cite{kim}. It would also be worth
exploring the same procedure for the general class of traversable
wormholes solutions derived in \cite{peter1}.

\subsubsection*{Acknowledgment}
One of us (MJ) would like to thank Farook Rahaman for sharing useful
comments to improve this work.

\end{document}